# Gate-tuning of graphene plasmons revealed by infrared nano-imaging

Z. Fei[1], A. S. Rodin[1], G. O. Andreev[1], W. Bao[2,3], A. S. McLeod[1], M. Wagner[1], L. M. Zhang[4], Z. Zhao[2], M. Thiemens[5], G. Dominguez[6], M. M. Fogler[1], A. H. Castro-Neto[7], C. N. Lau[2], F. Keilmann[8], D. N. Basov[1]

[1]Department of Physics, University of California, San Diego, La Jolla, California 92093, USA

[2]Department of Physics and Astronomy, University of California, Riverside, California 92521, USA

[3]Materials Research Science and Engineering Center, University of Maryland, College Park, Maryland 20742, USA

[4]Department of Physics, Boston University, Boston, Massachusetts 02215, USA

[5]Department of Chemistry and Biochemistry, University of California, San Diego, La Jolla, California 92093, USA

[6]Department of Physics, California State University, San Marcos, California 92096, USA

[7]Graphene Research Centre and Department of Physics, National University of Singapore, 117542, Singapore

[8]Max Planck Institute of Quantum Optics and Center for Nanoscience, 85714 Garching, Germany

**Surface plasmons are collective oscillations of electrons in metals or semiconductors enabling confinement and control of electromagnetic energy at subwavelength scales[1-5]. Rapid progress in plasmonics has largely relied on advances in device nano-fabrication[5-7], whereas less attention has been paid to the tunable properties of plasmonic media. One such medium—graphene—is amenable to convenient tuning of its electronic and optical properties with gate voltage[8-11]. Through infrared nano-imaging we explicitly show that common graphene/$SiO_2$/Si back-gated structures support propagating surface plasmons. The wavelength of graphene plasmons is of the order of 200 nm at technologically relevant infrared frequencies, and they can propagate several times this distance. We have succeeded in altering both the amplitude and wavelength of these plasmons by gate voltage. We investigated losses**

**in graphene using plasmon interferometry: by exploring real space profiles of plasmon standing waves formed between the tip of our nano-probe and edges of the samples. Plasmon dissipation quantified through this analysis is linked to the exotic electrodynamics of graphene[10]. Standard plasmonic figures of merits of our tunable graphene devices surpass that of common metal-based structures.**

In general, surface plasmons can exist in any material with mobile charge carriers whose response to electric field remains reactive, i.e., whose in-plane momentum- and frequency-dependent complex conductivity, $\sigma(q,\omega)=\sigma_1+i\sigma_2$, is predominantly imaginary. Of particular interest are plasmons with high momenta $q_p \gg \omega/c$, which may be utilized for extreme concentration of electromagnetic energy[1-5]. In conventional bulk metals, the frequencies of such high-$q$ plasmons reside in the visible or ultraviolet ranges. In graphene, they are expected to appear in the terahertz and infrared (IR) domains[12]. However, these high-$q$ IR plasmons are dormant in conventional spectroscopy of graphene. Here we utilized the scattering-type scanning near-field optical microscope (s-SNOM) to experimentally access high-$q$ plasmons by illuminating the sharp tip of an atomic force microscope (AFM) with a focused IR beam (Fig. 1a). The momenta imparted by the tip extend up to a few times $1/a$, where $a \approx 25$ nm is the curvature radius of the tip[13], thus spanning the typical range of IR plasmon momenta $q_p$ in graphene[12]. The spatial resolution of s-SNOM is also set by $a$, and proves to be an order of magnitude smaller than the plasmon wavelength $\lambda_p$. The direct observable of our method—the scattering amplitude $s(\omega)$—is a measure of the electric field strength inside the tip-sample nanogap. Consequently, the s-SNOM technique enables spectroscopy[13] and IR nano-imaging of graphene plasmons without the need to fabricate specialized periodic structures[14]. Our imaging data elucidate real-space characteristics of IR plasmons in graphene such as reflection, interference and damping. All these phenomena can be readily manipulated with gate voltage – a noteworthy property unattainable in metal-based plasmonics.

To probe directly the properties of graphene plasmons, we utilize a frequency $\omega = 892$ cm$^{-1}$ corresponding to the wavelength $\lambda_{IR} = 11.2$ μm in the IR regime where the plasmon is unimpeded by the surface optical phonon supported in graphene/SiO$_2$/Si structures[13]. The nano-imaging results are shown in Figs. 1b-e, where we plot normalized near-field amplitude $s(\omega) = s_3(\omega)/s_3^{Si}(\omega)$. Here, $s_3(\omega)$ and $s_3^{Si}(\omega)$ are the 3$^{rd}$ order demodulated harmonics of the near-field amplitude measured for the given sample and for a Si reference sample, respectively[13]. The near-field amplitude $s(\omega)$ tracks real-space variations in the local electric field underneath the tip, enabling exploration of surface phonon polaritons and surface plasmons[17,18].

In Fig. 1b, we present a $s(\omega)$ image acquired at the graphene-SiO$_2$ interface revealing periodic oscillations of the $s(\omega)$ signal extending along the graphene edge. Point and circular defects (Fig. 1d) trigger circular fringe patterns. Line defects (Fig. 1b) produce elongated, elliptical patterns. We observed fringes at both sides of the single/bilayer graphene boundary (Fig. 1c). Finally, strongly tapered corners of graphene (Fig. 1e) reveal the two groups of fringes oriented along both edges of graphene in the field of view. In all cases, the periodicity of the fringe patterns is around 100 nm.

Images in Fig. 1 are consistent with the following scenario. Illuminated by focused IR light, the AFM tip launches plasmon waves of wavelength $\lambda_p$ propagating radially outward from the tip. Sample edges or defects act as (imperfect) reflectors of the plasmon waves, directing them back to the tip. Therefore complex patterns of interference between launched and reflected plasmons should form inside graphene. We emphasize that our experimental technique does not capture instantaneous snapshots of these complex patterns. Instead, while the tip "launches" plasmon waves propagating in all directions, it only "detects" the cumulative near-field plasmonic signal arising underneath it. This stands in rough analogy with the operating principle of sonar echolocation. In Fig. 2a, we sketch the plasmon interference pattern in the form of plasmon amplitude revealing standing wave oscillations between the tip and sample edge. As the tip is scanned towards the edge, it registers these oscillations with periodicity given by $\lambda_p/2$ as shown in bottom panels of Fig. 2a. Our plasmon interference interpretation is supported by theoretical estimates of the wavelength $\lambda_p$. The plasmon dispersion of a two-dimensional (2D) metal residing at the interface of vacuum (dielectric constant $\varepsilon_0 = 1$) and a substrate with dielectric function $\varepsilon_{sub}(\omega)$ is given by the formula:

$$q_p = \frac{i\omega\kappa}{2\pi\sigma}, \quad \kappa(\omega) = \kappa_1 + i\kappa_2 \equiv \frac{\varepsilon_0 + \varepsilon_{sub}(\omega)}{2}. \tag{1}$$

Assuming that the conductivity $\sigma$ of graphene takes a Drude form with relaxation time $\tau$, one can rewrite equation (1) as:

$$q_p = \frac{\hbar^2 \kappa(\omega)}{2e^2 E_F} \omega(\omega + \frac{i}{\tau}). \tag{2}$$

Derivation of these equations and further refinements are discussed in the Supplementary Information. The real part of $q_p = q_1 + iq_2$ determines the plasmon wavelength $\lambda_p = 2\pi/q_1$ and the ratio between imaginary part and real part defines the plasmon damping rate $\gamma_p = q_2/q_1$. In graphene, the Dirac-like dispersion of the Fermi energy $E_F = \hbar v_F k_F$ with Fermi momentum $k_F = \sqrt{\pi|n|}$ [8], implies $|n|^{-1/2}$ scaling of the plasmon momentum with the carrier density $n$ at fixed $\omega$. Here $v_F \approx c/300 = 10^6$ m/s is the Fermi velocity. Finally, using frequency $\omega = 892$ cm$^{-1}$ and $n \approx 8\times10^{12}$ cm$^{-2}$ determined from the micro-Raman probe (see below) at the graphene edge, we find $\lambda_p \approx 200$ nm from equation (2), which is roughly twice the distance between fringes in Figs. 1b-e.

The images in Figs. 1b-e contain rich insights on processes governing plasmon propagation and losses on the surface of graphene. It is therefore instructive to examine line profiles along the direction normal to the sample edges. In Fig. 2b we show a plot obtained by averaging 150 such profiles—a procedure used to improve the signal-to-noise ratio. We find that the fringe widths increase from the interior to the edge of graphene, implying the plasmon wavelength $\lambda_p$ likewise increases. This behavior is due to enhancement of the carrier density $n$ near the sample edge, which is verified by our micro-Raman experiments (inset of Fig. 2b)[20,21]. Thus, plasmonic interference patterns reported in Fig. 2b uncover a utility of IR imaging for the nanoscale determination of local carrier density in graphene. In Fig. 2b we also show modeling results of plasmon profiles following a procedure detailed in the Supplementary Information. Our modeling provides a quantitative account of plasmon interferometry data. The carrier density profile (red curve in the inset of Fig. 2b) and the damping rate $\gamma_p$ constitute the adjustable parameters of the model. Since plasmons in our experiments are launched and detected by the same point source (AFM tip), the interference amplitude necessarily exhibits decay from the sample edge even when the damping rate is assumed to be vanishingly small (blue trace in Fig. 2b). The best fit to the amplitude profile is achieved for significantly stronger damping with $\gamma_p = 0.135$.

According to equations (1,2), the plasmon wavelength $\lambda_p$ is directly determined by the carrier density $n$. We experimentally demonstrate this unique aspect of graphene plasmonics through imaging under gate bias, displayed in Fig. 3a. Over a range of $V_g$ values from +30 V to -20 V, the hole density $n$ in our samples increases monotonically, a consequence of significant unintentional hole doping present even in ungated graphene/SiO$_2$/Si structures (inset of Fig. 2b). This tuning of carrier density produces systematic variations in the plasmonic profiles: fringe amplitude and periodicity are both enhanced with increasing $n$. By inferring $\lambda_p$ directly from the fringe width, we observe a systematic decrease in $\lambda_p$ with the reduction in hole density. Our gate-dependent data for $\lambda_p$ approximately follow the $\lambda_p \propto |n|^{1/2}$ law predicted for monolayer graphene[22]. In contrast, the plasmon damping rate does not show clear gate dependence and is roughly equal to $0.135 \pm 0.015$ at all $V_g$. This magnitude of $\gamma_p$ significantly exceeds theoretical estimates for graphene with similar electronic mobility $\mu \approx 8000$ cm$^2$/V·s [12].

It is important to understand why plasmon damping in our structures is abnormally strong. According to equations (1,2) two additive contributions define damping rate as $\gamma_p \approx (\sigma_1/\sigma_2) + (\kappa_2/\kappa_1)$. The first term is associated with plasmonic losses implicit to graphene, whereas the second term describes losses due to the SiO$_2$ substrate. At $\omega = 892$ cm$^{-1}$, we estimate $\kappa = 2.52 + 0.13i$ and hence, $\kappa_2/\kappa_1 \approx 0.05$, based on our ellipsometric measurements of SiO$_2$/Si wafers. The resulting value of $\sigma_1/\sigma_2 \approx 0.08$ is unexpectedly high, three to four times higher than the estimate of $\sigma_1/\sigma_2 = (\omega\tau)^{-1}$ one obtains from equation (2) using the relaxation rate $\tau^{-1} \sim 20$ cm$^{-1}$, corresponding to a typical DC mobility of our samples. This discrepancy affords two possible interpretations. Excessive losses may originate from an enhanced electronic relaxation rate at infrared frequencies compared to that established in DC transport. Alternatively, losses may be unrelated to free carrier mobility/dynamics and may instead be associated with extrinsic factors such as surface irregularities. Our plasmonic interferometry data in Figs. 2,3 provide strong support for the former hypothesis. Indeed, these images yield $\gamma_p$ and $\lambda_p$ (Fig. 3b) and thus allow us to determine the complex optical conductivity of graphene (inset of Fig. 3b) based on the formulae: $\sigma_2 \approx \frac{\omega\kappa}{4\pi^2}\lambda_p$; $\sigma_1 = \sigma_2(\gamma_p - \kappa_2/\kappa_1)$. We remark that these relations between plasmonic parameters ($\lambda_p$, $\gamma_p$) and the complex optical conductivity $\sigma$ of graphene hold true for any plasmonic material for which $\sigma_1 << \sigma_2$. Therefore these expressions apply even if the frequency-dependent conductivity deviates from the simple Drude model. The appeal of this analysis lies in establishing a link between real-space plasmonic profiles and the optical constants inferred from conventional IR spectroscopy. Our imaging data reveal that the real part of the conductivity of graphene is as high as $\sigma_1 \approx 0.5 e^2/h$. This value greatly exceeds the theoretical estimate of $\sigma_1(\omega = 892 \text{ cm}^{-1}, n = 8 \times 10^{12} \text{ cm}^{-2}, \tau^{-1} = 20 \text{ cm}^{-1}) = 0.13 e^2/h$ obtained from a model of non-interacting Dirac quasiparticles weakly scattered by disorder[23-25], but is comparable to IR spectroscopy results for back-gated graphene on $SiO_2/Si$[10]. In an ideal, non-interacting graphene $\sigma_1(\omega < 2E_F)$ is vanishingly small due to the phenomenon of "Pauli blocking"[10]. Thus the source of strong plasmonic losses in our back-gated samples is traced to the unexpectedly large magnitude of $\sigma_1$. This result supports the notion of prominent many-body effects in graphene beyond the picture of non-interacting Dirac fermions[27-29]. Further experiments on suspended graphene as well as devices employing various types of dielectric substrates such as BN are needed to disentangle the roles of electron-electron and graphene-substrate interactions in the dissipation we observe at infrared frequencies. Our work uncovers an experimental path and analysis methodology for these future studies of many-body interactions in graphene.

Infrared nano-imaging experiments reported here established graphene/$SiO_2$/Si structures as a potent plasmonic medium that enables voltage control of both the plasmon wavelength and amplitude. Higher gate voltages than used in our study will allow for the binary on/off switching of plasmon propagation with a possibility of potentially local control by a top gate or a biased tip. The plasmon wavelength in graphene $\lambda_p \approx 200$ nm is one of the shortest imaged for any material whereas the propagation length is on par with Au in experiments monitoring strongly confined plasmons launched by AFM tips[18].

An important figure of merit $\lambda_{IR}/\lambda_p$ = 50-60 for our back-gated devices surpasses that of conventional Ag-based structures[12]. Intrinsic plasmonic losses in graphene that we analyzed in detail can be substantially reduced or even eliminated through population inversion[30]. We stress that plasmon tuning is realized here in the architecture of a Metal(graphene)-Oxide-Semiconductor(Si) device: a ubiquitous system in modern information processing. Furthermore, the performance of even the first generation of plasmonic devices reported here and in Ref. 15 is rather promising against non-tunable metal-based structures[12]. For all these reasons we believe that graphene may be an ideal medium for active infrared plasmonics.

**Methods Summary**

**Infrared nano-imaging using s-SNOM:** Our s-SNOM apparatus (Neaspec) is based on an atomic force microscope (AFM) operating in tapping-mode. Measurements were performed at an AFM tapping frequency of $\Omega$ = 270 kHz and tapping amplitude of 40 nm. The shaft of the conducting AFM tip acts as an antenna that boosts the efficiency of near-field interaction[26]. The back-scattered signal registered by the detector is strongly dependent on the tip-sample distance. This enables isolation of the genuine near-field contribution from the overall back-scattered signal, which is periodically modulated at harmonics of the tapping frequency $\Omega$. The $m^{th}$ harmonic component of this signal $s_m e^{i\phi_m}$, termed the $m^{th}$ demodulated signal (here $m$ = 3), represents the desired near-field contribution.

**Samples and devices:** Our graphene samples were obtained by mechanical cleavage of bulk graphite and then transferred to $SiO_2$/Si substrate. To avoid surface contamination by lithographic procedures, bulk graphite connected to our graphene sample was utilized as an electrode in our back-gating experiments. To verify the gating functionality of our devices, we first performed a spectroscopic study of the hybrid plasmon-phonon resonance at various gate voltages and found good agreement with the published data[13]. Transport measurements of graphene samples fabricated following identical procedures indicate that the typical mobility of our graphene samples is about 8000 $cm^2$/Vs. Plasmon imaging experiments were completed for more than 30 graphene samples. All these structures exhibited highly reproducible behavior and consistent values for the plasmon wavelength and damping. The data displayed in Figs.1-3 were obtained for devices with some of the weakest damping revealing the largest number of plasmonic oscillations. Nevertheless, even in these devices the plasmonic losses are stronger than expectations based on typical electronic mobility measurements.

**Micro-Raman measurements:** According to previous studies, the G-peak position of the graphene Raman signal is directly linked to its carrier density[19-21]. Therefore, the G-peak profile shown in the inset of Fig. 2b reflects the range of the variation in graphene carrier density close to the edge[20-22]. Our micro-Raman experiments were carried out using a Renishaw inVia Raman microscope equipped with a $50\times$, NA = 0.75 long-distance objective, a 1800 l/mm grating and an XY stage with the resolution of 100 nm. The spot size in these experiments is limited by diffraction. Therefore, the fragment of the line profile of the G-peak frequency shown in the inset of Fig. 2b is instrumentally broadened.

**Acknowledgements**

Authors acknowledge support from AFOSR, and ONR. The analysis of plasmonic losses and many body effects is supported by DOE-BES. W.B, Z.Z, and C.N.L were supported by NSF DMR/1106358, ONR N00014-09-1-0724, ONR/DMEA H94003-10-2-1003 and FENA focus center. G.D. and M.T. were supported by NASA. M.F. is supported by UCOP and NSF PHY11-25915. A.H.C.N. acknowledges NRF-CRP grant R-144-000-295-281. M.W. thanks the Alexander von Humboldt Foundation for financial support. F.K. was supported by Deutsche Forschungsgemeinschaft through the Cluster of Excellence Munich Centre for Advanced Photonics.

**Author contributions**

All authors were involved in designing the research, performing the research, and writing the paper.

**Additional information**

The authors declare no competing financial interests; F.K. is cofounder of Neaspec, producer of the s-SNOM apparatus used in this study. Supplementary information accompanies this paper on www.nature.com/naturematerials. Reprints and permission information is available online at http://npg.nature.com/reprintsandpermissions. Correspondence and requests for materials should be addressed to D. N. B.**References**

1. Atwater, H. A. The promise of plasmonics. *Sci. Am.* **296**, 56–63 (2008).
2. West, P. R. et al. Searching for better plasmonic materials. *Laser & Photon. Rev*. 1-13 (2010).
3. Stockman, M. I. Nanoplasmonics: The physics behind the applications. *Phys. Today* **64**, 39-44 (2011).
4. Maier, S. A. Plasmonics: Fundamentals and Applications Ch. 4 (Springer, New York, 2007).
5. Schuller, J. A. et al. Plasmonics for extreme light concentration and manipulation. *Nature Mater.* **9**, 193-204 (2010).
6. Nagpal, P., Lindquist, N. C., Oh, S.-H. & Norris, D. J. Ultrasmooth patterned metals for plasmonics and metamaterials. *Science* **325**, 594-597 (2009).
7. Lal, S., Link, S. & Halas, N. J. Nano-optics from sensing to waveguiding. *Nature Photon.* **1**, 641-648 (2007).
8. Castro-Neto, A. H., Guinea, F., Peres, N. M. R., Novoselov, K. S. & Geim, A. K. The electronic properties of graphene. *Rev. Mod. Phys.* **81**, 109-162 (2009).
9. Wang, F. et al. Gate-variable optical transitions in graphene. *Science* **320**, 206-209 (2008)
10. Li, Z. Q. et al. Dirac charge dynamics in graphene by infrared spectroscopy. *Nature Phys*. **4**, 532-535 (2008).
11. Vakil, A. & Engheta, N. Transformation optics using graphene. *Science* **332**, 1291-1294 (2011).
12. Jablan, M., Buljan, H. & Soljacic, M. Plasmonics in graphene at infrared frequencies. *Phys. Rev. B* **80**, 245435 (2009).
13. Fei, Z. et al. Infrared nanoscopy of Dirac plasmons at the graphene-$SiO_2$ interface. *Nano Lett.* **11**, 4701-4705 (2011).
14. Ju, L. et al. Graphene plasmonics for tunable terahertz metamaterials. *Nature Nano.* **6**, 630-634 (2011).

**Figure Legends**

**Figure 1 | Infrared nano-imaging experiment and results. a,** Schematic of an infrared nano-imaging experiment at the surface of graphene (labeled as G) on $SiO_2$. Green and blue arrows display the directions of incident and back-scattered light, respectively. Concentric red circles illustrate plasmon waves launched by the illuminated tip. **b-e,** Images of infrared amplitude $s(\omega = 892\text{ cm}^{-1})$ defined in the text taken at zero gate voltage. These images show a characteristic interference pattern close to graphene edges (blue dashed lines), defects (green dashed lines), at the boundary between single (G) and bilayer (BG) graphene (white dashed line). Additional features marked with the arrows in **e** are analyzed in Refs. 15,16. Locations of boundaries and defects were determined from AFM topography taken simultaneously with the near-field data. Scale bars are 100 nm in all panels. All data were acquired at ambient conditions.

**Figure 2 | Spatial variation of the near-field amplitude at the graphene edge.** For all panels, graphene extends at $L > 0$, and $SiO_2$ not covered by graphene is displayed at $L < 0$. **a,** Illustration of interference between tip-launched plasmon waves (white) and their reflection (green) from the edge at $L = 0$. Solid and dashed lines correspond to positive and negative field maxima of the propagating plasmon, respectively. False color plots of the absolute value of electric field $|E_z|$ reveal standing waves formed between the tip and the edge. Left and right panels show snapshots of destructive (minimum signal) and constructive (maximum signal) interference underneath the tip, respectively. Scale bar, $0.5\lambda_p$. We also plot profiles of $|E_z|$ underneath the tip versus its distance to the edge. The blue circles and arrows mark the positions of the tip. **b,** Experimental (grey) and calculated (color) $s(\omega)$ line profiles at zero gate voltage. Inset shows the G-peak positions

inferred from micro-Raman data and the carrier density profile (red line) we used to model the plasmonic standing wave (panel **b**). The Raman G-peak positions are associated with the variation of the local carrier density in graphene (right-hand scale)[19-21].

**Figure 3 | Electrostatically tunable plasmon in back-gated graphene. a,** $s(\omega)$ line profiles perpendicular to the graphene edge at various gate voltages. Inset illustrates our gate bias setup. **b**, Gate-dependent plasmon wavelength $\lambda_p$. Inset displays the real and imaginary part of the optical conductivity of graphene at various gate voltages estimated from $\lambda_p$ and $\gamma_p$ as described in the text. These conductivity data directly show that the response of graphene is predominantly reactive: $\sigma_2 \gg \sigma_1$ thus fulfilling an essential precondition for excitation of plasmons.

Figure 1

Figure 2

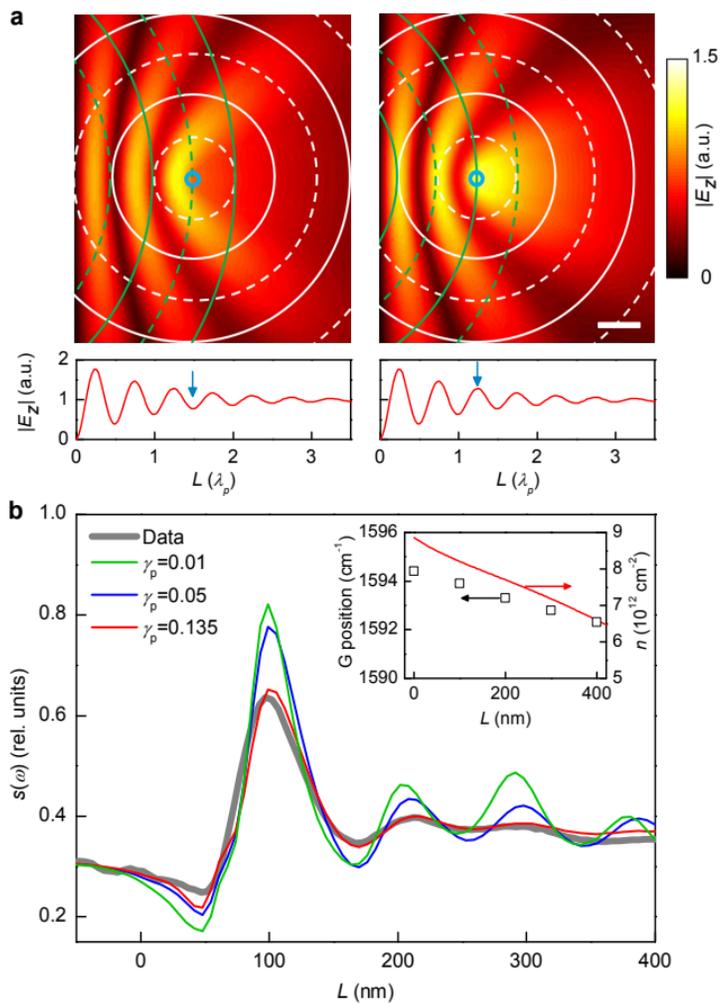

Figure 3

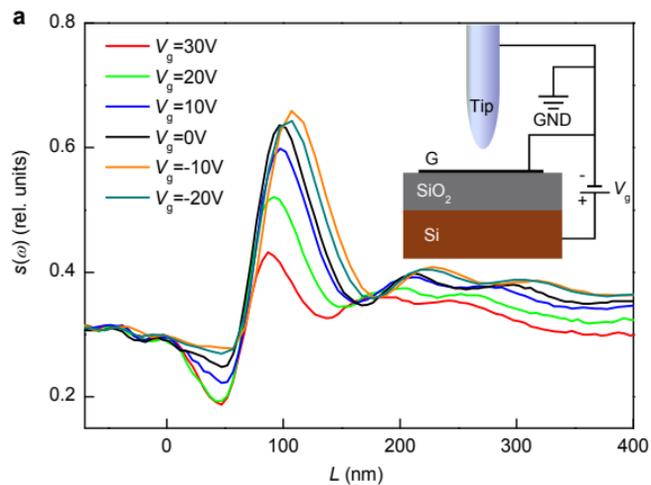

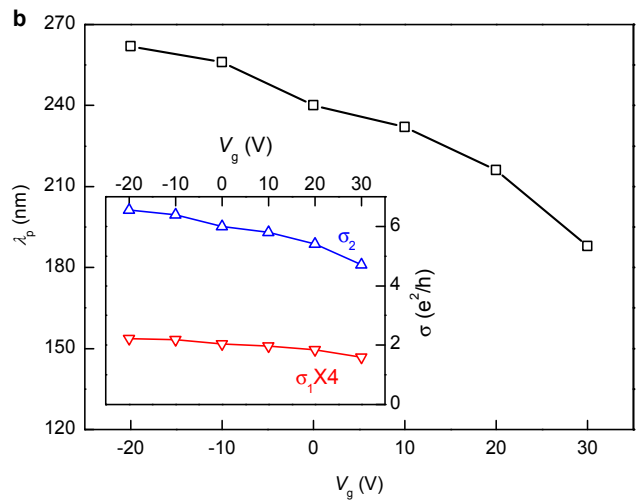